\documentclass[10pt,conference]{IEEEtran}
\IEEEoverridecommandlockouts

\usepackage{cite}
\usepackage{amsmath,amssymb,amsfonts}
\usepackage{algorithmic}
\usepackage{graphicx}
\usepackage{textcomp}
\usepackage{xcolor}
\def\BibTeX{{\rm B\kern-.05em{\sc i\kern-.025em b}\kern-.08em
    T\kern-.1667em\lower.7ex\hbox{E}\kern-.125emX}}

\usepackage{pbalance}
\usepackage[hyphens]{url}

\makeatletter
\newcommand{\linebreakand}{%
  \end{@IEEEauthorhalign}
  \hfill\mbox{}\par
  \mbox{}\hfill\begin{@IEEEauthorhalign}
}
\makeatother

\begin{document}
\title{An Approach for a Supporting Multi-LLM System for Automated Certification Based on the German IT-Grundschutz\\
}

\author{%
\IEEEauthorblockN{Muth, Lea}
\IEEEauthorblockA{\textit{Department of Mathematics and Computer Science} \\
\textit{Freie Universität Berlin}\\
Berlin, Germany \\
Lea.Muth@fu-berlin.de}
\and
\IEEEauthorblockN{Margraf, Marian}
\IEEEauthorblockA{\textit{Department of Mathematics and Computer Science} \\
\textit{Freie Universität Berlin}\\
Berlin, Germany \\
Marian.Margraf@fu-berlin.de}
}

\maketitle

\begingroup
\renewcommand\thefootnote{}
\footnotetext{\textcopyright~2025 IEEE. Personal use of this material is permitted. Permission from IEEE must be obtained for all other uses, in any current or future media, including reprinting/republishing this material for advertising or promotional purposes, creating new collective works, for resale or redistribution to servers or lists, or reuse of any copyrighted component of this work in other works.}
\endgroup

\begin{abstract}
This paper presents a novel approach to perform semi-automated BSI IT-Grundschutz certification using a Multi-Large Language Model system (MLS) with Hybrid Retrieval-Augmented Generation (HybridRAG). 
Facing the challenges of the Network and Information Security Directive 2 (NIS2) directive, a shortage of specialists, and high implementation costs, our MLS architecture aims to increase efficiency, reduce costs, and support certifiers in maintaining the quality of security concepts while meeting the increased demand for certifications of newly affected companies.
The system combines Large Language Models (LLMs) and Knowledge Graphs (KGs) to support different phases of the certification process, including protection needs assessment, modeling, IT-Grundschutz check, measure consolidation, and subsequent realization.

Our architecture addresses the growing demand for security concepts and offers an approach to handle the digital security challenges introduced by NIS2.
\end{abstract}
\section{Introduction}
The NIS2~\cite{NIS2} represents a significant expansion of cybersecurity (CS) regulations across the European Union (EU), with implementation deadlines fast approaching. As of October 18, 2024, affected organizations must ensure NIS2 conformity, with the directive becoming fully mandatory from March 2025. Medium-sized companies (more than 50 employees or over €10 million turnover p.a.) must implement comprehensive security concepts starting in March 2025. This directive mandates the establishment of robust security frameworks, emergency plans, and the capability to report security incidents within 24 hours of discovery~\cite{NIS2}. 
Non-compliance carries severe penalties, up to €10 million or 2\% of global turnover p.a. (whichever is higher) for essential entities, and up to €7 million or 1.4\% for important entities~\cite{NIS2}.

%
The directive expands the scope of affected entities in Germany from approximately 4,500 to around 29,500 - 1,800 being operators of critical infrastructure~\cite{bmi_1}. Some companies work with classified information needing a higher level of protection~\cite{bmi_1}. Those organizations come with widely varying security requirements and require different security levels. The creation of their security concept requires considerable expertise, as well as continuous maintenance, adaptation, and review. 

%
This leads to the next problem: The acute shortage of qualified certifiers capable of producing and validating the demanded security concepts. The European Network and Information Security Agency reports that the percentage of IT Full-Time Equivalents dedicated to information security has continued to decline from 11.9\% to 11.1\%, with 32\% of organizations and 59\% of Small and Medium Enterprises (SMEs) struggling to fill CS roles. This trend is particularly noteworthy and critical, as 89\% of companies expect to need additional human resources to comply with NIS2 and beyond for CS in general~\cite{enisa}. 

%
Furthermore, the financial burden of certifying security concepts is prohibitive for many organizations, with implementation costs to meet NIS2 regulations estimated at €31.2 billion p.a. across the EU~\cite{front_1}. Implementing the NIS2 Directive in Germany is estimated to increase costs for businesses by up to €7.3 billion~\cite{front_2}. The cost of implementing the directive varies considerably between different sectors and company sizes, with SMEs bearing a proportionately greater burden in relation to their turnover. SMEs will have to invest in new security technologies and build up the necessary expertise by training existing staff or hiring new specialists. All of this must be done with the context of a general shortage of security experts and the threat of severe penalties~\cite{front_1, front_2}. Emergency plans must be regularly reviewed and updated to address new threats, with organizations required to conduct at least annual reviews and simulated exercises. This implies significant resources and expertise that many SMEs can hardly provide.

This paper presents an MLS architecture based on AI and the IT-Grundschutz Compendium (IT-GC) of the Federal Office for Information Security (BSI) to help address the critical challenges posed by the NIS2 Directive. 
In Germany, the BSI IT-Grundschutz ('Fundamental Security' or 'Basic Protection') methodology is widely recognized as a framework. Certifiers use IT-GC to develop comprehensive security concepts~\cite{grundschutz_kompendium, front_2}, therefore, we have adopted this established standard as the basis for our MLS architecture. The IT-GC provides a systematic approach to establishing and operating an Information Security Management System (ISMS). It offers detailed guidance on developing security policies, selecting appropriate security requirements, and implementing security measures in practice. Although IT-Grundschutz provides a solid baseline with its standard protection, companies must implement additional measures to ensure full compliance with NIS2, as stated by BSI~\cite{bmi_2}.

We present a modular architecture that utilizes multiple LLMs and Knowledge Graph Databases (KG-DBs) to semi-automate processes within the certification process. 
Our MLS architecture automates repetitive and time-consuming work steps within the certification process through our system, giving certifiers more time to process additional security concepts. The modular structure allows for individual components to be replaced with organization-specific components as required by the varying organizational security requirements.
\section{Background }
The IT-Grundschutz methodology ~\cite{grundschutz_standards} developed by BSI provides a systematic approach to establishing and maintaining information security in organizations. The methodology is compatible with the international standard ISO/IEC 27001. The BSI Standards~~\cite{grundschutz_standards} (200-1, 200-2, 200-3, and 200-4) and the IT-GC~\cite{grundschutz_kompendium} together form the foundation of the IT-Grundschutz methodology.
%

The \textit{BSI Standard 200-1 Information Security Management Systems} contains general requirements for an ISMS, guidelines for creating security processes, and security concepts. It establishes the terminology and basic principles guiding the security process (200-1, ~\cite{grundschutz_standards}).
%
The \textit{BSI Standard 200-2 IT-Grundschutz Methodology }contains the basic protection methodology, the audit basis for certification, and detailed specifications for the design, implementation, and enhancement of a compliant management system. It includes an explanation of the individual steps to create a security concept. This standard can be adapted to the requirements of organizations of various types and sizes, providing a flexible framework tailored to an organization's specific environment (200-2, ~\cite{grundschutz_standards}).
%
The \textit{BSI Standard 200-3 Risk Analysis} based on IT-Grundschutz contains procedures for risk analysis and guidelines for conducting independent risk assessments for objects with high protection requirements or for which no suitable basic protection module exists. This risk analysis procedure is based on elementary threats and consists of specific steps, including risk identification, risk classification, risk treatment, and consolidation of the security concept (200-3, ~\cite{grundschutz_standards}).
%
The \textit{BSI Standard 200-4 Business Continuity Management} specifies the requirements of the emergency management module. It focuses on Business Continuity Management as a holistic process to minimize IT operational disruptions (200-4, ~\cite{grundschutz_standards}).
%

The IT-GC follows a modular approach to allow better structuring and planning in the complex area of information security and its security assessment. It contains modules organized in different thematic layers, providing concrete security measures for IT systems and processes~\cite{grundschutz_kompendium}. 
Each module describes generic requirements within the respective layer that must be adapted to the specific framework conditions of an organization. These modules are regularly updated, and the requirements formulated in the modules describe the general measures that should be implemented to achieve adequate security measures for a given level of protection (200-2, ~\cite{grundschutz_standards}).

The methodology is designed to make the creation of security concepts simple and efficient. In contrast to traditional risk analysis, in which threats and vulnerabilities have to be identified individually, the modules already contain ready-to-use risk analyses~\cite{grundschutz_kompendium}.
This aims to reduce the analytical effort of performing a gap analysis between the security requirements of the relevant modules and the safeguards already implemented within the organization~\cite{grundschutz_kompendium}.

Depending on their protection requirements, companies can choose between three types of protection: Basic Protection, Standard Protection, and Core Protection. 
Organizations that meet the Standard Protection requirements can obtain a certification for a respective scope (Informationsverbund). This paper focuses on this type of protection for the certification.
\section{Certification based on IT-Grundschutz}
A certification based on IT-Grundschutz for a Standard Protection goes through nine stages, as illustrated in Fig.~\ref{fig}:
\begin{figure}[htbp]
\centerline{\includegraphics[width=0.36\textwidth]{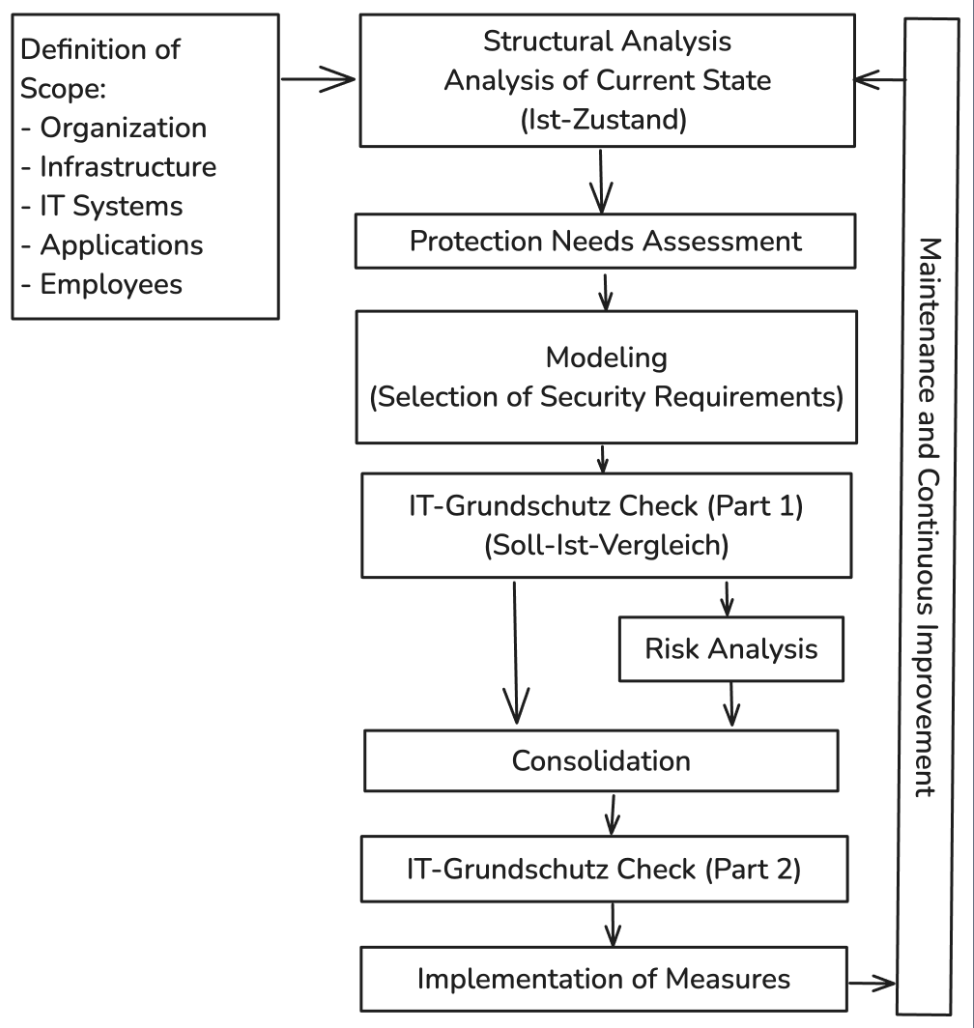}}
\caption{Nine-Step Process for IT-Grundschutz Certification under Standard Protection. The diagram shows the repetitive nine-step process for carrying out certification under ISO 27001 based on IT-Grundschutz.}
\label{fig}
\end{figure}
\begin{enumerate}
    \item \textbf{Definition of the Scope of Application}: The scope defines the boundaries of the ISMS and determines which assets, processes, departments, and activities are covered by the ISMS. 
    \item \textbf{Structural Analysis}: The structural analysis records all components of an information network and their interactions within the defined scope. It captures the current state to determine necessary protective measures. This analysis involves four steps: documenting business processes and applications, creating a network plan, surveying IT, Industrial Control Systems, and Internet of Things systems, and identifying spaces. Components are recorded as objects and can be grouped to reduce complexity. For each object, the required protective measures are identified. 
    \item \textbf{Protection Needs Assessment}: 
    The protection needs assessment determines the confidentiality, integrity, and availability (CIA) requirements for each object. IT-Grundschutz categorizes these needs as "Normal", "High", or "Very high", with detailed risk analysis required for "High" and above. Protection needs are inherited following principles like the maximum principle, e.g., if an application is classified as "Very high"; this protection need is then inherited by the running server and the allocated server room. Inheritance considers cumulative effects, where multiple "Normal" applications on one system may elevate its overall protection need.
    \item \textbf{Modeling}: 
    During modeling, the information network is replicated using the standardized modules from the IT-GC. For common components such as "Windows clients", corresponding modules exist. However, there is not a suitable module for every specialized system, therefore, objects that cannot be modeled with the existing modules must undergo an individual risk analysis. 
    \item \textbf{IT-Grundschutz Check}: 
    The IT-Grundschutz Check (IT-G-Check) is a comparison that evaluates the degree of implementation of the requirements from the IT-GC, which are derived from the modules, and the actual degree of implementation within the company. Implementation is documented as "yes", "partially", "no", or "dispensable". This process yields a To-do list of measures needed to meet the requirements. Repeated application of the IT-G-Check (Parts 1 \& 2, see Fig.~\ref{fig}) re-examines implemented measures to ensure that new security requirements are also addressed.
    \item \textbf{Risk Analysis}: 
    The standard requirements of IT-Grundschutz are designed for objects with normal protection needs and enable security without a separate risk analysis. An explicit risk analysis is still necessary for objects with increased protection needs ("High" and "Very High"), for which no suitable modules exist, or that deviate from the standard assumptions of IT-Grundschutz due to special operating conditions and are used differently than intended in the IT-GC. In these cases, the risk analysis examines the additional security requirements. 
    \item \textbf{Consolidation}: 
    During consolidation, measures are reviewed for conflicts and overlap. This ensures coherent security measures without contradictory or redundant elements.
    \item \textbf{Implementation of the measures}:
     A detailed realization plan is drawn up for all security measures that have not yet been fully implemented. This plan includes careful budgeting of the necessary resources, prioritization of the measures according to their urgency and effectiveness, binding implementation deadlines, and clear assignment of responsibilities to responsible employees or teams. 
    \item \textbf{Maintenance and Improvement}: 
    Security is an iterative process; an ISMS must continue to be maintained, continuously updated, and its effectiveness verified even after the certification process is completed.
\end{enumerate}
\section{Related Work}
In ~\cite{KG_CS_ref}, the authors provide a systematic overview of the applications of CS KGs in various phases of the comprehensive CS lifecycle that organizations follow to protect their networks and systems, encompassing prevention, detection, and response activities, including situation awareness, threat discovery, and attack investigation. Specific applications are presented that support operators and managers in decision-making, operations, vulnerability management, malware attribution, and analysis. The authors conclude their graph-based approach enables security analysts to visualize attack paths and potential vulnerabilities across network infrastructure, correlate seemingly unrelated security events to identify coordinated attacks, apply reasoning and inference capabilities to discover hidden patterns and predict potential threats, integrate heterogeneous security data from multiple sources into a unified knowledge model, and support automated reasoning about security situations to reduce cognitive load on analysts.
This work provides a broad foundation for understanding the utility of KGs in CS, which our architecture leverages for the specific task of IT-Grundschutz certification. While they cover a wider range of applications, we focus on automating a specific, regulation-driven process. 
%

In ~\cite{KG_ICS_ref}, a framework is presented that is based on a KG and integrates previous work on Products, Processes, and Resources (PPR), as well as cause-effect modeling. Bayesian networks are used to estimate failure probabilities and propagate these through the KG to identify possible attack chains and their impacts on safety issues. The authors specifically work on developing a standards-based model using Resource Description Framework and Web Ontology Language to construct KGs that represent safety and security-relevant information in production environments. Their approach combines the PPR Model with Failure Cause-Effect relationships. This integration allows for a comprehensive analysis of security vulnerabilities and their potential safety implications and reveals the relationships between various elements in their system.
This work is relevant because it demonstrates the use of KGs for representing complex relationships in a safety and security context, similar to how our architecture represents relationships between IT-Grundschutz components. However, we focus on compliance with a specific standard, whereas they focus on a broader safety and security analysis in production systems.

The author of ~\cite{LLM_compliance_ref} focused on developing an automated methodology for classifying legal provisions in the field of food safety and improving compliance checks of regulatory artifacts using LLMs. The author compared various LLMs (BERT, GPT-3.5) and baselines (LSTM, keyword search) in terms of accuracy and efficiency. A three-step process was used for compliance checking including content chunking, prompt construction, and LLM-based compliance checking. The results showed that LLMs can outperform traditional methods in terms of time, cost, and accuracy in classifying legal provisions. For compliance checking of Data Processing Agreements with the General Data Protection Regulation, the author found significant improvements when moving from sentence-level to paragraph-level analysis. 
This work is directly relevant as it explores the application of LLMs to regulatory compliance, which is the core focus of our architecture. We build upon this by combining LLMs with a KG (HybridRAG) to improve accuracy, reduce hallucinations, and adapt to the IT-Grundschutz standard. 
%

\cite{LLM_compliance_3_ref} examined LLMs' ability to transform regulatory texts into actionable tasks in finance. Their case study of a global bank using for transaction monitoring and regulatory updates achieved a 60\% reduction in manual hours spent on compliance reviews and experienced a 30\% drop in compliance costs within the first year. The study also found that the bank enhanced detection accuracy, leading to a 50\% decrease in flagged false positives. 
These results indicate that LLM-based systems improve efficiency and accuracy in complex compliance tasks and real-world applications. Our proposed MLS architecture builds upon this, aiming to achieve comparable benefits within IT-Grundschutz certifications. Our design prioritizes generalizability, allowing for potential adaptation to other compliance frameworks beyond IT-Grundschutz.
\section{Proposed Method: Multi-LLM System}
\subsection{BSI's example of work - Recplast GmbH}
The BSI provides a working example of IT-Grundschutz certification using a fictional company. This example includes essential documents such as the security guideline, the guideline for risk analysis, the guideline for the control of documents and records, the guideline for internal ISMS auditing, the guideline for the control of corrective and preventive measures, and an IT structure analysis (IT-SA). 
This example serves as the blueprint for our MLS architecture. 
\subsection{Multi-LLM architecture with HybridRAG}
\begin{figure}[t!]
\centering
\includegraphics[width=0.46\textwidth]{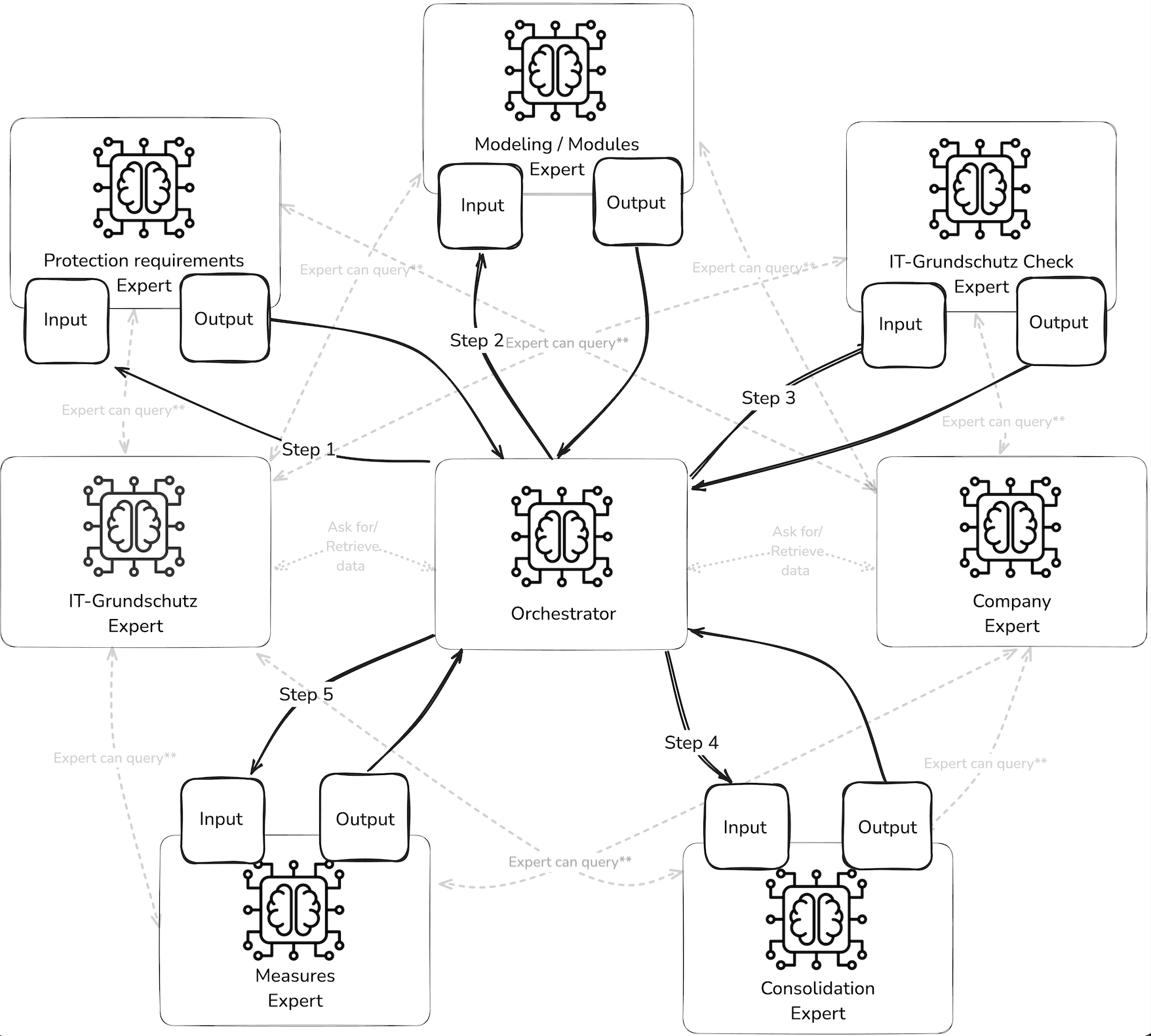}
\caption{Multi-LLM architecture with HybridRAG. Our MLS is based on eight LLM experts; Two knowledge experts (IT-Grundschutz expert, Company expert), five task experts, and one orchestrator. The task experts each have a predefined processing sequence and a defined input and output (boxes). The orchestrator controls the scheduling between the other experts and validates their inputs and outputs. The knowledge experts can be consulted by all other experts (dotted lines) to obtain information.}
\label{fig_multiLLM}
\end{figure}

We propose the combination of multiple task-specific LLMs and HybridRAGs for our semi-automated IT-Grundschutz certification. The MLS architecture leverages the varied model capabilities to automate the certification process and aims to reduce manual effort, as illustrated in Fig.~\ref{fig_multiLLM}. 

Several criteria led to this type of MLS architecture;
%
A HybridRAG~\cite{hybridRAG} combines the strengths of graph-based and vector-based retrieval methods. Information is stored as text fragments and their corresponding vector embedding and saved in structured graphs with entities (nodes) and their relationships (edges).
HybridRAG uses vector-search to identify semantically similar vectors while simultaneously employing graph structures to capture relational information. Through this combination, HybridRAG offers both global context through vector-based representations and detailed local context via graph-based relationships~\cite{GraphRAG_LLM_hallus}.
%
A selection criterion for HybridRAG was that it outperforms GraphRAG and VectorRAG systems in terms of answer accuracy and information relevance for corporate documents in the finance sector~\cite{hybridRAG}. Finance data and compliance documents share significant overlap as both adhere to strict regulatory frameworks and legal standards, both contain sensitive information requiring careful handling and redaction, and both must follow specific formatting and content standards to meet industry and legal requirements. Financial documents often contain complex tables, graphs, and multi-modal elements, making AI processing challenging. Understanding such documents may also require connecting information across distant sections.

This is particularly important for our use case of certification, as a HybridRAG architecture can recognize and utilize connections between different fragments of information that are not directly connected within the text~\cite{GraphRAG_LLM_hallus, graphRAG_OG}.
%

We chose task-specific LLMs as experts because they can handle large text corpora and adapt to the certification process requirements~\cite{llm_OG}.
%
LLMs can break down complex certification questions, e.g., concerning various aspects of IT-Grundschutz, into multiple subquestions, and thus potentially increase the accuracy of responses~\cite{chain_of_verification}. This ability to decompose queries is particularly valuable, as certification questions often address several areas simultaneously. 
%

Beyond that, they can receive complex natural language queries about IT security and transform them into graph queries. This allows certifiers to interact with the system in their familiar technical language without knowing any graph query language, making them the human-in-the-loop experts. This eliminates the need for additional IT experts and increases user-friendliness, as no in-depth AI knowledge is required~\cite{GraphRAG_LLM_hallus_2, graphRAG_OG}. The models can identify and specifically query relevant entities and relationships in KGs from contextual questions.
%

Since LLM hallucinations remain a persistent issue, the LLMs' responses require continuous verification. Our MLS architecture includes the following;
%
Specific metrics are indispensable for the continuous evaluation of statement quality. The \textit{Faithfulness metric} measures how precise LLM responses align with reference documents~\cite{GraphRAG_LLM_hallus}, which is particularly important when verifying compliance with IT-GC. \textit{Answer Relevancy} assesses whether the output answers the query informatively and precisely~\cite{GraphRAG_LLM_hallus}, which is crucial to evaluating the relevance of security requirements. \textit{Precision} and \textit{Recall} measure accuracy and completeness in identifying relevant security requirements~\cite{LLM_compliance_ref} and are particularly valuable for assessing the IT-Grundschutz modeling phase. Prompt Alignment\cite{Wu2023PromptAlign} determines whether an LLM follows the instructions of the input prompt, ensuring adherence to the guidelines. \textit{BERTScore} evaluates the semantic similarity between the LLMs' output and reference texts and is excellently suited for comparison with expert assessments\cite{BERTScore}. The \textit{Hallucination Score} identifies false or fabricated information in the LLM output~\cite{GraphRAG_LLM_hallus} and is crucial for quality assurance in security recommendations.
%
The introduced metrics require an evaluation dataset, which needs to be revised by an expert. For the IT-Grundschutz KG this could be done by German authorities. For company specifics, a security expert generates the test set through a human-in-the-loop approach during the certification process.
%
Combining these metrics with regular spot checks creates a robust quality assurance system. \textit{Contextual Relevance} measures how relevant the retrieved context is to the user's query or input\cite{GraphRAG_LLM_hallus}. Using LLM-as-a-judge methods, these metrics can be applied in an automated and scalable manner, where an LLM evaluates the answers of another LLM based on predefined criteria~\cite{graphRAG_OG}.
Due to the high level of automation, the option allows for an automated continuous monitoring of the model's behavior to detect deviations or unexpected patterns that could indicate potential security risks without waiting for the next expert review. The selection of the specific models for the LLM-as-a-judge task should take into account that models can be self-preference biased. For example, GPT-4 exhibits a significant degree of self-preference bias, tending to favor its own outputs over those from other models\cite{self_preference_bias}.
%

The use of HybridRAG in conjunction with task-specific LLMs has proven to be particularly effective in minimizing hallucinations, as LLM responses based on a structured knowledge base significantly increase the likelihood of finding relevant information, thus reducing speculative information~\cite{GraphRAG_LLM_hallus, GraphRAG_LLM_hallus_2}. 
HybridRAG achieves a 6\% reduction in hallucinations compared to GraphRAG ~\cite{GraphRAG_LLM_hallus, graphRAG_OG} on regulatory documents, including the Digital Operational Resilience Act from the EU with the Federal Financial Institutions Examination Council guidelines from the United States, resembeling our use-case. 
HybridRAG can represent the complex relationships between IT-Grundschutz modules, requirements, and measures in the form of structured entities and their relationships, serving as the source of information for the LLM \cite{KG_CS_ref, GraphRAG_LLM_hallus}. 
The semantic structure of a graph enables context-related information provision, tailored to specific certification tasks.
The IT-Grundschutz is available in machine-readable format and can be transferred into a KG (200-1,~\cite{grundschutz_standards}). The hierarchical and modular nature of the IT-Grundschutz, with its defined relationships between modules, requirements, and measures, is particularly well-suited to representation in a KG, which helps to ground the LLMs' responses in factual information and reduce the likelihood of hallucinations.
Particularly valuable is HybridRAG's ability to identify cross-references between different areas of IT-Grundschutz. For example, if a module depends on several others, the graph can explicitly represent these relationships and make them usable for queries.
%
The increased transparency of decision-making is another advantage of HybridRAG systems and LLMs, allowing inspection of data source connections and relationships within the KG, enhancing transparency, auditability, and trustworthiness~\cite{GraphRAG_LLM_hallus} — which are essential qualities for auditing AI-driven certifications.
%

HybridRAG supports continuous updates of the underlying KG, as IT-Grundschutz is regularly updated. This flexibility is important when integrating new regulatory requirements, as the compliance landscape is continuously evolving.
%

Further, it is important to highlight the increase in efficiency, as HybridRAG systems combined with LLMs have more efficient processing compared to traditional RAG. HybridRAG requires up to 80\% fewer tokens for LLM response generation, which improves the scalability of the system~\cite{GraphRAG_LLM_hallus, graphRAG_OG}.
%

Moreover, the described benefits can be applied to represent the organization. A company’s IT infrastructure, processes, responsibilities, and security measures can be modeled as a KG. This representation links regulatory requirements to relevant company areas, enabling gap analysis and generating context-specific certification recommendations for each organization. This results in another decisive advantage: interoperability between different KGs, e.g., a company KG that includes information about specific software versions used. By merging the KG with the IT-Grundschutz KG, the system could identify vulnerabilities specific to those versions and recommend targeted patches, going beyond the IT-Grundschutz recommendations. Merging KGs is not trivial. It requires resolving ontological mismatches, entity disambiguation, and semantic conflicts while maintaining referential integrity across heterogeneous data structures.

%
In this paper, we focus on identifying the most labor-intensive parts of the methodology that constitute “diligent work” to achieve quick and efficient assistance with repetitive, time-consuming processes. Our MLS architecture follows the iterative schema of the certification illustrated in Fig.~\ref{fig}.

The Orchestrator guides the overall certification process, activating task-specific experts (task experts) sequentially and managing the data flow between steps. To ensure access to authoritative information, task experts can directly consult the knowledge experts as needed, leveraging their HybridRAG interface to retrieve relevant context from the IT-Grundschutz and company KGs. To implement the specific function of each task expert, they are configured through prompt engineering, including assigning a specialized persona, providing detailed task instructions, defining the expected input data structure (including retrieved context), and specifying the required output format (e.g., JSON schema) to ensure consistency and validation by the Orchestrator.

The \textit{orchestrating expert} manages the scheduling process, delegating tasks to other experts with diverse assignments and skills, and verifies the in- and output of the five task experts. 
Another duty of the orchestrator is to ensure correctness by validating and refining the answers of the task experts. This is achieved by validating the output against predefined JSON schemas and regular expressions that match expected answer formats. Combining these measures with regular spot checks creates a robust quality assurance system \cite{chain_of_verification}.

The \textit{knowledge experts} (IT-Grundschutz Expert, Company Expert) utilizing KGs provide all task experts and the orchestrator with a common, consistent knowledge base through the knowledge experts. This reduces contradictory statements and improves coherence throughout the certification process.
%

The \textit{protection requirements expert} is provided with the dependencies between the objects, the IT-SA, the protection requirements, and the protection requirements categories. 
First, it extracts the dependency relationships between the objects and then applies the inheritance rules systematically. It considers the maximum principle for vertical inheritance, the cumulative effects of multiple applications on a system, and the distribution effects in redundant systems. Afterward, it identifies cross-dependencies between applications that would not be captured by purely vertical inheritance. Finally, it verifies all dependencies. The KG's structure supports the completeness of the considered dependencies and enables a thorough examination of inheritance rules, leading to more reliable and comprehensive management of protection requirements~\cite{KG_CS_ref}.
It returns all protection requirements and the reasons for the classifications, the inheritance chains showing the transfer of requirements, and a list of critical assets with high protection requirements with higher priority.
The expert can significantly accelerate the time-consuming manual assignment and calculation of protection requirements for complex information networks with numerous dependencies.

The \textit{modeling expert} is given the IT-SA and the protection requirements for each object (1.). 
The expert first performs the structural analysis on the IT-SA, identifying all objects (servers, clients, network components, etc.). Subsequently, it assigns the objects to suitable modules of the IT-GC, making correct assignments even with different terminology, thanks to their semantic capabilities. Afterward, it identifies modeling gaps, including the recognition of target objects for which no suitable standard modules exist, and marks them for a later individual risk analysis.
As a result, it returns a structured assignment table (objects to modules), including reasons for module selection, the list of non-modelable objects, and a list of dependency relationships between modules.
This automation significantly relieves security experts, as the manual assignment of objects to modules is a highly time-consuming and labor-intensive process. The automation potential is also supported by the recent developments of the BSI, which aim at machine-readable documentation and a higher degree of automation\footnote{\url{https://www.bsi.bund.de/SharedDocs/Downloads/DE/BSI/Veranstaltungen/Grundschutz/1GS_Tag_2025/Aktuelle_Entwicklungen_Ausblick_IT-GS.pdf}, (Accessed: 2025-02-28) }.
%

The \textit{IT-Grundschutz Check Expert} (3.) is handed the protection requirements and the security levels (1., 2.), as well as access to all company documents, including guidelines. 
It extracts relevant requirements from the assigned modules, creating structured checklists that consider basic, standard, and increased requirements based on protection levels. Through document analysis, the expert can compare company policies and configurations with these requirements, identifying content matches despite terminology differences. Then, for each requirement, it proposes an implementation status, identifies documentation gaps, and prioritizes action needs based on criticality, thus creating the basis for an effective realization plan. Lastly, it enables continuous monitoring through updates when changes occur in the IT-GC or company documents and can identify and incorporate new or modified requirements.
The expert returns a catalog of the current implementation statuses for each requirement. Including a list of identified opportunities for improvement, a prioritized list of measures including reasons for the assessments, and a comparison with previous checks identifying progress or new gaps.
This automation saves considerable time, especially in follow-up audits, improves the quality and consistency of the assessment, and enables the recognition of indirect evidence for meeting requirements through language understanding.
%

The \textit{Consolidation expert} (4.) is given the results (1.), (2.), of the IT-G-Check (3.), potential individual risk analyses, and a list of available resources, budget restrictions and organizational framework conditions.
The expert starts its task by identifying semantically similar and functionally overlapping measures from different modules. For example, different requirements for access rights in various modules could lead to conflicts. It then analyzes those dependencies between measures and can suggest an optimized sequence of implementation that leverages synergies and reduces overall effort. Afterward, it consolidates different terminologies for similar concepts, identifies potential gaps in the security architecture, and conducts cost-benefit analyses to optimize resource allocation. It can make suggestions for optimizing resource allocation by identifying measures that simultaneously fulfill multiple requirements with minimal effort. 
As a result, it creates a structured, prioritized realization plan free of redundancies and contradictions, including justifications for decisions.
By analyzing the explicit entirety of all measures, the expert can identify potential gaps in the security architecture not covered by the modules. Particularly valuable is the expert's ability to go beyond mere word equality and recognize functional equivalences between differently formulated measures, thereby uncovering more subtle redundancies and contradictions, which enables a more profound consolidation and significantly relieves the security expert while simultaneously improving the quality and consistency of the realization plan.

The \textit{Measures expert} (5.) is handed the consolidated action plan cleared of redundancies and contradictions (4.), the IT-G-Check list with implementation statuses and prioritization criteria including specifications (3.), as well as full access to all company documents.
The expert starts by prioritizing measures based on protection requirements, risk potential, and implementation effort. It then creates a detailed, context-specific implementation proposal tailored to the company. This proposal includes technical configuration steps, organizational measures, structured realization plans, and documentation to verify successful implementation. This facilitates audits and improves long-term maintainability. Further, the expert can create specific training materials for measures that require employee training, saving costs for external courses. 
It then returns a prioritized implementation plan, including a sequence of actions, detailed implementation instructions, a responsibility matrix, test plans and success criteria, standardized templates for documenting, and, if needed, target group-specific training materials.
Implementing the proposed measures is the core of actual security improvements. Although the actual implementation requires human expertise, LLMs can significantly support and optimize the process.

\subsection{Restrictions of our architecture}
The quality and reliability of our MLS for supporting certification processes according to IT-Grundschutz must be considered in a differentiated manner. Such a system can achieve significant efficiency gains; however, certain limitations must be taken into account:
%
Neither AI systems nor human experts can perform error-free work. Human certifiers are subject to fatigue, limited capacities (e.g., time), and make mistakes in complex or repetitive tasks. An MLS can partially compensate for these weaknesses by working consistently and processing large amounts of data without losing concentration. Nevertheless, such a system also does not produce 100\% correct statements.
Therefore, our MLS is designed as a support tool, not as a replacement for human expertise. The final responsibility and decision-making authority must remain with qualified experts who can validate and contextually classify the results generated by the system. This human-machine collaboration leverages the strengths of both sides: the efficiency and scalability of the AI system, and the judgment and experience of human experts. However, it should be remarked that humans tend to have overconfidence in the models, and model outputs can be deceivingly confident, although false.
\section{Initial Evaluation}
Taking the BSI's example as benchmark data, the first implementation of the rudimentary functionalities of the Protection Needs Assessment showed promising results. For our implementation, we used LangChains' Graph Transformer library to process the unstructured IT-Grundschutz documents (PDFs and BSI data) and transform them into structured, graph-based representations. For a fast initial implementation, we have opted for GPT-4o-mini whenever we used an LLM. Any other model with equivalent performance can be used instead. Later, it is intended to use local models to perform the tasks for which we currently use GPT-4o-mini, as some data - especially company data - cannot be uploaded or shared with model providers for privacy reasons.

An LLM (GPT-4o-mini, text-embedding-3-small, recursive character text splitter, chunk size of 1024, and an overlap of 200) was applied to extract entities and relationships for the KGs. 
We then fed both extracted graph-based documents into a Neo4j KG-DB and combined it with Neo4js' Hybrid Retriever to create our knowledge experts. This retriever enables HybridRAG, combining vector-based semantic similarity search with graph traversal based on the KG. 
 
Given that the certification process is a step-wise procedure, and as a complete automatization is envisioned, we began extracting the protection requirements from the IT-Grundschutz. As standards could be changed, this is the first step to automatization. We did so by querying the IT-Grundschutz expert to generate a complete list of all layers, including all modules and their respective requirements for Basic Protection, Standard Protection, and Core Protection.

All ten layers and all 111 modules were found by our model. The requirements were extracted correctly by 92.2\%. Saving those responses and asking the model again led to a 94.6\% retrieval rate. The visual inspection in the KG showed no abnormalities and complete data, this suggests that the prompting could be optimized.

To spot-check the second KG-DB, the company's knowledge expert was asked to find the company's current protection level for a given business process for each of the CIA categories. The answer retrieval was at 92.1\%. It should be emphasized that after simply asking the model to double-check the results, the retrieval went to 95.7\%. This indicates a possible improvement in the query parameters and not a mistake in the underlying KG-DB.

A first rudimentary part of the modeling expert was implemented. As ground truth the BSI's sample data containing a module assignment was used. The expert was asked to suggest the top-n matching modules for each of the 87 target objects with a given protection requirements definition. If the correct module was within the suggested list of n-modules, this was considered to be correct. For the top-5, the expert provided a 86.21\% correct suggestion; for the top-3, the expert provided a 83.90\% correct suggestion; and for the top-1, the expert provided a 74.71\% correct suggestion.

These initial tests were only used to see if the data was extracted correctly from the provided documents. No optimization or fine-tuning was done, so the authors emphasize that this is a first small result and should not be considered a full evaluation, but simply a motivation for continuing the implementation. A complete evaluation can only be done once the entire MLS has been implemented, which has not yet happened at this stage, and when authentic company data is available. The datasets to perform such a comprehensive evaluation are currently being collected.
 
\section{Future Research}
The presented MLS architecture supports the automation of IT-Grundschutz certification and offers a promising approach to address current challenges: the shortage of personnel among certifiers, rising costs, growing demand for security concepts, and the continuous updating of the still young compliance requirements. Through intelligent automation, certifiers can be significantly relieved, costs can be substantially reduced, and compliance with IT-Grundschutz standards can be made considerably more efficient.
To fully exploit the potential of this technology and provide even more comprehensive support to certifiers, expanding the automation approach to the entire certification process would be of great interest. The authors plan to further research and implement the conceptual idea of this paper in a prototype system for a full evaluation.

\end{document}